\documentclass[aps,pre,reprint,floatfix,nofootinbib,superscriptaddress]{revtex4-1}
\pdfoutput=1
\usepackage{graphicx}
\usepackage{amssymb,amsfonts,amsmath}
\usepackage{color}
\usepackage{mathtools}
\usepackage{todonotes}
\usepackage{grffile}
\usepackage{gensymb}

\newcommand{\vort}{\zeta}
\newcommand{\phimelt}{\phi_\text{melt}}
\newcommand{\phij}{\phi_\text{J}}
\newcommand{\phicc}{\phi_{C_3\text{-}C_2}}
\newcommand{\psiafm}{\psi_\text{AFM}}
\newcommand{\psih}{\psi_\text{H}}

\begin{document} 

\title{Spatiotemporal order and emergent edge currents in active spinner materials}
\author{Benjamin C. van Zuiden}
\affiliation{Instituut-Lorentz, Universiteit Leiden, 2300 RA Leiden, The Netherlands}
\author{Jayson Paulose}
\affiliation{Instituut-Lorentz, Universiteit Leiden, 2300 RA Leiden, The Netherlands}
\author{William T. M. Irvine}
\affiliation{James Franck and Enrico Fermi Institutes, Department of Physics, The University of Chicago, Chicago, IL 60637}
\author{Denis Bartolo}
\affiliation{Univ. Lyon, ENS de Lyon, Univ. Claude Bernard, CNRS, 
Laboratoire de Physique, F-69342 Lyon, France}
\author{Vincenzo Vitelli}
\affiliation{Instituut-Lorentz, Universiteit Leiden, 2300 RA Leiden, The Netherlands}

\begin{abstract}
  Collections of interacting, self-propelled particles have been extensively
  studied as minimal models of many living and synthetic systems from bird
  flocks to active colloids. However, the influence of active rotations in the
  absence of self-propulsion (\emph{i.e.} spinning without walking) remains less
  explored. Here, we numerically and theoretically investigate the behaviour of
  ensembles of self-spinning dimers. We find that geometric frustration of dimer
  rotation by interactions yields spatiotemporal order and active melting with
  no equilibrium counterparts. At low density, the spinning
  dimers self-assemble into a triangular lattice with their orientations
  phase-locked into spatially periodic phases. The phase-locked patterns form
  dynamical analogues of the ground states of various spin models, transitioning
  from the 3-state Potts antiferromagnet at low densities to the striped
  herringbone phase of planar quadrupoles at higher densities. As the density is
  raised further, the competition between active rotations and interactions
  leads to melting of the active spinner crystal. Emergent edge currents, whose
  direction is set by the chirality of the active spinning, arise
  as a non-equilibrium signature of the transition to the active spinner liquid
  and vanish when the system eventually undergoes kinetic arrest at very high
  densities. Our findings may be realized in systems ranging from liquid crystal
  and colloidal experiments to tabletop realizations using macroscopic chiral
  grains.\end{abstract}

\maketitle

The last two decades have seen significant progress in our understanding
of active matter. Early theoretical
progress\cite{Marchetti_review,Cavagna_review,Vicsek_review} has been
accompanied by the engineering of soft materials made of self-propelled
polymers, colloids, emulsions, and
grains~\cite{Dogic2013,Bausch2010,Bricard2013,Palacci2013,
  Ginot2014,Thutupalli2011,Deseigne2010,Kudrolli2010}, which exhibit novel
nonequilibrium phenomena. 
Prominent examples include phase separation of
repulsive spheres, giant number fluctuations away from criticality, and
long-range orientational order in two-dimensional
flocks~\cite{TonerTu,Cates_review,Toner_review}.

The systems mentioned above have in common the characteristic that constituents
acquire translational momentum due to active propulsion, but rotate only in
response to collisions or diffusion. By contrast, insights into the consequences
of active \emph{rotation} without self-propulsion remain scarce, even though
this situation is relevant to a wide range of experimental
systems~\cite{Snezhko2016}
including spinning microorganisms~\cite{Drescher2009,Petroff2015}, treadmilling
proteins~\cite{Denk2016}, sperm-cell and microtubule
aggregates~\cite{Riedel2005,Sumino2102}, shaken chiral grains~\cite{Tsai2005},
light-powered chiral colloids~\cite{Saglimbeni2015}, thermally and chemically
powered liquid crystals~\cite{Tabe2003,Oswald2015}, electrorheological
fluids~\cite{Lemaire2008}, and biological and synthetic cilia driven by rotary
molecular motors~\cite{Furthauer2013a}.

Until now, theoretical and numerical studies on ensembles of active spinners
have separately addressed their phase dynamics and their spatial organization.
The emergence and robustness of synchronized rotation in lattices of
hydrodynamically-coupled rotors~\cite{Uchida2010,Uchida2010a} has been studied
as an archetype of Kuramoto dynamics in coupled oscillator
systems~\cite{Acebron2005}. In these models the lattice geometry is imposed, a
situation relevant for instance to the propagation of metachronal waves at the
surface of ciliated tissues~\cite{Nonaka2002,Guirao2010,Button2012,Brumley2015}.
Local orientational synchronization has also been observed in self-organized
disordered arrays of rotating rods~\cite{Kirchhoff2005,Kaiser2013}.
A separate class of numerical studies has
been devoted to the {\em spatial structures} of ensembles of active spinners
interacting either via contact or hydrodynamic
interactions~\cite{Lenz2003,Nguyen2014,Sabrina2015,Spellings2015,Yeo2015,Goto2015,Aragones2016}.
Special attention has been paid to phase separation in binary mixtures of
counter rotating spinners and to hydrodynamic interactions yielding spatial
ordering.

Here, we bridge the gap between these two lines of research. Combining numerical
simulations and analytical theory we demonstrate the inherent interplay between
the spatial structure and the phase dynamics of active spinners. We uncover a
generic competition between monopole-like interactions that dominate at large
separations, and shorter range multipole gear-like interactions. We find that
their interplay frustrates ordered states but also yields novel spatiotemporal
order and unanticipated collective flows including edge currents.

We study a prototypical system of soft dimers interacting \emph{via} repulsive
interactions and undergoing unidirectional active rotation as sketched in
Fig.~1. When isolated, dimers spin in response to the active
torque, attaining a steady-state spinning speed due to background friction. As
they get closer, the multipole character of the pair interactions resists the
rotation of adjacent dimers (Fig.~1b--c). At very high densities,
the relative motion of neighbours is completely obstructed
(Fig.~1d).
By tuning the density, we explore how the frustration between monopole and multipole interactions
plays out as their relative strengths are varied (Movie S1). We observe transitions from
collections of independently spinning dimers to unusual crystal states which are
ordered in particle position as well as orientation over time (Movies S2 and S3),
to active spinner liquids, to jammed states.
Repulsive interactions with boundaries also obstruct
spinning (Fig.~1e); to compensate, the system channels the
rotational drive into linear momentum, giving rise to robust edge currents and
collective motion (Movie S1). 

Our model system consists of a 2D ensemble of $N$ like-charge dimers,
each consisting of two point particles of mass $m$ connected by a stiff link
of length $d$, Fig.~1a. Point particles interact only \emph{via}
a repulsive pair potential of the Yukawa form $b
e^{-\kappa r}/r$, where $b$ sets
the overall strength of the repulsion, $r$ is the inter-particle separation
distance, and $\kappa$ is the inverse screening length, see
Fig.~1. By setting $\kappa^{-1} \sim d$, we discourage dimer links
from crossing each other and also
maximize the orientational dependence of the effective pair interaction
between dimers. 

Each dimer is actively driven implemented by a 
torque $\tau = F d$ implemented as a force dipole (Fig.~1a).
Energy is dissipated by drag forces
acting on each particle with associated drag
coefficient $\gamma$. The equations of motion for the position, ${\bf r}_i$ and
orientation $\theta_i$ of the $i^{\rm th}$ dimer are
\begin{align}
    2m\mathbf{\ddot{r}}_i &=   -2\gamma \mathbf{\dot{r}}_i-\partial_{{\mathbf r}_i}\sum_{j\neq i} {V}({\mathbf r}_j-{\mathbf r}_i,\theta_i,\theta_j)  
    \label{eq:eqnofmotion}\\
    I{\ddot \theta}_i&=\tau-{\gamma_\Omega }{\dot \theta}_i -\partial_{{\theta}_i} \sum_{j\neq i}{\cal V}({\mathbf r}_j-{\mathbf r}_i,\theta_i,\theta_j)
    \label{eq:eqnofmotion2}
\end{align}
where $I=md^2/2$ and $\gamma_\Omega = \gamma d^2/2$ are the moment of inertia
and rotational friction coefficients respectively, and the position- and orientation-dependent
interaction potentials $V$ and $\cal V$ are derived from the Yukawa pair
interactions between the point particles. An isolated dimer attains a steady state of counterclockwise
rotation about its center with a constant spinning speed $\Omega_0 =
\tau/\gamma_\Omega$ (Fig.~1b). In contrast to systems where the dimer
orientation is slaved to an external field (e.g., colloids
driven by a rotating magnetic field~\cite{coq2011,Yan2015}), the instantaneous
dimer orientation is not dictated by the internal drive in our system.

\begin{figure}[t]
  \centering
  \includegraphics{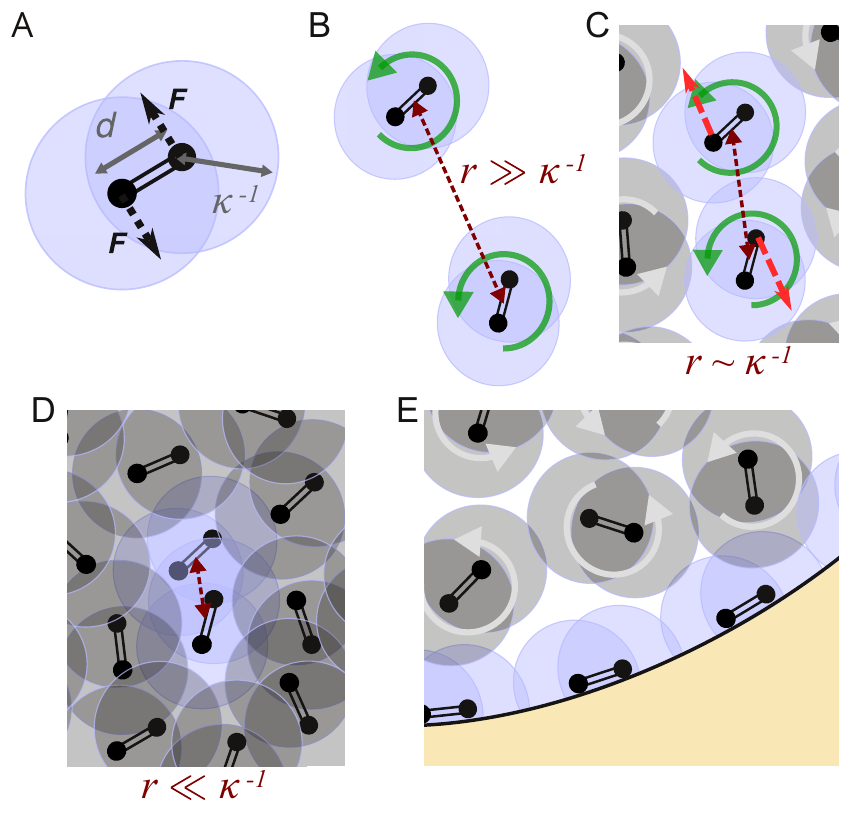}
  \caption{{\bf Competing rotation and interactions in active spinners.}
    {\bf a,} Make-up of a single self-spinning dimer,
    consisting of a pair of identically-charged particles (black dots), connected by a rigid
    rod of length $d$ (double line). Particles repel each other with a Yukawa interaction
    with  screening length $\kappa^{-1}$ which determines the soft exclusion zone
    (light blue discs), beyond which the repulsion falls off exponentially with
    distance. Each particle experiences a force of 
    magnitude $F$ and direction indicated by dotted arrows, oriented to provide
    zero net force and a net torque $\tau = Fd$ on the dimer at all times.
    {\bf b--d,} The density determines the influence of
    interactions on dimer dynamics. At large separations ({\bf b}), interactions are
    negligible and dimers freely rotate at the terminal angular velocity set by
    the activity and the background drag. As separations become comparable to
    the screening length ({\bf c}), adjacent dimers still rotate past each other
    but experience interaction forces (red dashed arrows show instantaneous
    force due to the interaction between two of the particles) that depend on their instantaneous
    orientations. At very high densities ({\bf d}), interactions completely obstruct dimer
    rotation.
    {\bf e,} Hard boundaries also obstruct dimer rotation, and their effect is
    transmitted into interior dimers by interactions.
  }
  \label{fig:intro}
\end{figure}

Upon rescaling distances by $\kappa^{-1}$ and time by $\Omega_0^{-1}$, the
dynamical equations are characterized by three dimensionless quantities:
$\kappa d$, $\alpha \equiv I\tau/\gamma_\Omega^2$ which measures the
characteristic dissipation time for angular momentum in
units of the spinning period, and $\beta^{-1} \equiv \tau/\kappa b$ which quantifies
the drive in units of the characteristic interaction energy scale.
We focus here on the competition between
rotational drive and interactions as the dimer \emph{density} is varied for fixed
$\alpha$ and $\beta$, as sketched in Fig.~1b--d. We constrain
ourselves to the asymptotic limit where both $\alpha \gg 1$ and $\beta \gg 1$. 

\section{Phase behavior}
We characterized the bulk behaviour of interacting spinners through simulations
under periodic boundary conditions in which the dimer density was varied by
changing the dimensions of the simulation box with constant screening parameter
$\kappa = 0.725/d$, particle number $N=768$, and activity parameters
$\alpha=131$ and $\beta = 133$, see Methods. Density is quantified by the
packing fraction $\phi = A \rho$, where $\rho$ is the number density of dimers
and $A = \pi (d+2\kappa^{-1})^2/4$ is the soft excluded area of a spinning dimer
on time scales $t \gg 1/\Omega_0$.
Fig.~2 characterises the phase behavior of our system \emph{via} 
changes in particle ordering, orientational ordering and dynamics in the
nonequilibrium steady states reached at long times. Nearly identical behavior is
observed for simulations with $N=3072$, indicating that finite-size effects are
negligible (Supplementary Figure S1). 

\subsection{Active spinner crystals}
At low packing fractions, the dimers self-organize into a hexagonal crystalline
pattern, with little or no change in position, as shown for two representative
densities in the first two panels of Fig.~2a. In this regime, the
repulsions between dimers give rise to a Wigner-like crystal, quantified by high
values of the bond-orientational order parameter $|\langle \psi_6 \rangle|$
(Fig.~2d, triangles). Although the dimers are highly restricted
in their position, they continue to spin without hindrance, attaining the same
angular speed as an isolated dimer ($ \langle \dot \theta \rangle \equiv \Omega \approx \Omega_0$,
Fig.~2e). Apart from small fluctuations, the orientation of dimer
$i$ at time $t$ has the form $\theta_i(t) = \Omega_0 t +\delta_i$ with the
angular phase $\delta_i$ defined up to a global phase shift. This state is
reminiscent of plastic crystals, but with the equilibrium fluctuations
of the orientational degrees of freedom replaced by active rotation: we term
this state an \emph{active spinner crystal}.

The crystals display ordering not only in dimer positions, but also in dimer
orientations which are phase-locked into regular spatial patterns (Movie S2 and
Movie S3). The angular phases
$\delta_i$ take on a few discrete values determined by the lattice
position. We find evidence for two distinct configurations. At low densities,
$\delta_i$ acquires one of three values $\{0,\pi/3,2\pi/3\}$, with no two
neighbors sharing the same value (Fig.~2c, first panel). This
pattern is identical to the equilibrium ground state of the 3-state Potts antiferromagnet
(3P-AFM) on the triangular lattice~\cite{Schick1977}. When $\phi > \phicc\approx
1.2$, the rotational symmetry of the pattern changes from $C_3$ to $C_2$ as
stripes of alternating $\delta_i \in \{0, \pi/2\}$ form along a
spontaneously-chosen lattice direction (Fig.~2c, second panel).
This phase is a dynamical analogue of the striped herringbone (H) phase observed
in lattices of elongated molecules~\cite{Mouritsen1982}. Local order parameters
$\psiafm$ and $\psih$ (defined in Methods) measure the extent to which phase
differences among neighbouring dimers match those prescribed by the respective
ordered states. As shown in Fig.~2f, the 3P-AFM and H states are
each observed over a range of densities.

To understand the origin of the phase-locked patterns, we study a minimal model
of the dimer-dimer interactions. To lowest order in dimer size $d$, each dimer
is a superposition of a charge monopole and a charge quadrupole. The monopole
repulsion arranges the dimer centres into a triangular crystal with lattice
constant $a \sim 1/\sqrt{\phi}$. We assume that the dimer positions are thus
fixed and focus on the orientation dynamics, \eqref{eq:eqnofmotion2}, due to the
quadrupolar interactions.  When averaged over the common rotation period $2\pi/\Omega_0$,
\eqref{eq:eqnofmotion2} reduces to $\partial_{\theta_i}\langle \sum_{j\neq
  i}{\cal V}({\mathbf r}_j-{\mathbf r}_i, \theta_i,\theta_j) \rangle_t=0$; \emph{i.e.}
the nonequilibrium steady states extremize the time-averaged potential energy
as a function of orientation.

Upon ignoring fluctuations around the constant-speed evolution $\theta_i(t)
= \Omega_0 t + \delta_i$, and considering only nearest-neighbour interactions
among dimers, the average effective energy takes the compact form
\begin{equation}
  \label{eq:avecyclepot}
  V_\text{eff} \equiv \langle \sum_{j\neq i}{\cal V}(\theta_i-\theta_j) \rangle_t = \sum_{\langle ij \rangle} \left[A_1 + A_2
  \left(\frac{d}{a}\right)^4 \cos 2(\delta_i - \delta_j)\right], 
\end{equation}
where $A_1$ and $A_2$ vary with density (see SI Text for details). For an
infinite lattice of dimers, $V_\text{eff}$ has arbitrarily many extrema.
However, the extrema can be exhaustively listed for a \emph{triangle} of
neighbouring dimers. Up to a global phase shift and vertex permutations, the
effective energy as a function of the phase shifts
$\{\delta_1,\delta_2,\delta_3\}$ on the triangle vertices has three unique
extrema at $\{0,\pi/3,2\pi/3\}$, $\{0,0,\pi/2\}$, and $\{0,0,0\}$. The 3P-AFM
and H phases respectively extend the first and second of these extrema onto the
infinite triangular lattice, and are thus also extremal states of the periodic
crystal. In fact, the 3P-AFM state is the \emph{global} energy minimum for
$V_\text{eff}$, as seen by mapping the effective energy to the antiferromagnetic
$XY$ model on the triangular lattice~\cite{Lee1984}.\footnote{The effective energy
  inherits a discrete and a continuous ground-state degeneracy from the
  antiferromagnetic $XY$ model. An arbitrary
  global phase shift gives the same state, but this is equivalent to a choice of
  $t=0$ in the description of the orientations. The discrete degeneracy is in
  the chirality of phase order ($0 \to \pi/3 \to 2\pi/3$ vs. $0 \to 2\pi/3 \to
  \pi/3$) upon circling a plaquette. Adjacent plaquettes always have opposite
  chirality, and the two possible chirality arrangements on the triangular
  lattice provide two distinct ground states.} The extremum with phase values
$\delta_i=0$, which would correspond to all dimers sharing the same orientation
at all times, maximizes the frustration of spinning by interactions and is not
observed in our simulations.

In summary, spinning dimers are frustrated. The spatiotemporal-crystal
states that are compatible with the mutual frustration of the position and
orientation degrees of freedom are captured by the extrema of the effective
potential $V_\text{eff}$. However, in principle, active spinner crystals could
harbour a multitude of other phase-locked patterns, which cannot be reduced to
repetitions of a single triangular unit but nevertheless extremize
$V_\text{eff}$. These may be accessible by modifying the initial or boundary
conditions, or the dynamics of approaching the nonequilibrium steady state.

\subsection{Melting and kinetic arrest}
We now elucidate how synchronized spinning motion frustrates positional order
and melts dense spinner crystals. As the packing fraction is increased, we
observe a \emph{loss} of crystalline ordering,  
signalled by a sharp drop in $|\langle \psi_6
\rangle|$ from 1 to $0.2$ at $\phi =\phimelt \approx 1.9$. This drop coincides
with the onset of diffusive dynamics of the dimer centres of mass at long times
(Supplementary Figure~S2). The diffusivity $D \equiv \lim_{t \to
  \infty} \langle |\mathbf{r}_i(t_0+t)-\mathbf{r}_i(t)|^2 \rangle_i/t$ is
nonzero for a range of densities above $\phimelt$, characteristic of a liquid
phase. Melting is accompanied by a disruption of the
phase-locked spinning dynamics, as quantified by (i) a drop in the average spin
velocity to below $0.1\Omega_0$ (Fig.~2e), (ii) a marked increase
in spin speed fluctuations, (Supplementary Figure~S3) and (iii)
a loss of H order in the orientations (Fig.~2f).
Fig.~2a--c (third column) shows a typical liquid configuration
with no discernible order in the positions, orientations, or spinning speeds.

The melting of the dimer crystal upon increasing the density, at odds with the
typical behaviour of athermal or equilibrium repulsive particles, is a direct
result of the orientational dependence of dimer-dimer interactions coupled with
the active spinning. The monopole part of the pair interaction is responsible
for the crystalline arrangement of dimer centres. The quadrupolar component generates
a gearing effect, which hinders the activity-driven co-rotation of adjacent dimers as shown
schematically in Fig.~1c. The competition between interactions and
active spinning results in geometrical frustration of the crystalline order, akin to the
frustration of antiferromagnetic Ising spins on the triangular lattice.
Increasing the density strengthens the quadrupolar component of the interactions
relative to the monopole component, 
\emph{de}stabilizing the crystal at the threshold packing fraction $\phimelt$. In the
liquid state, the frustration of in-place dimer rotation by interactions is
partially relieved by dimers constantly sliding past each other, at the cost of
crystalline and phase-locked order. 

\begin{figure*}[t]
  \centering
  \includegraphics{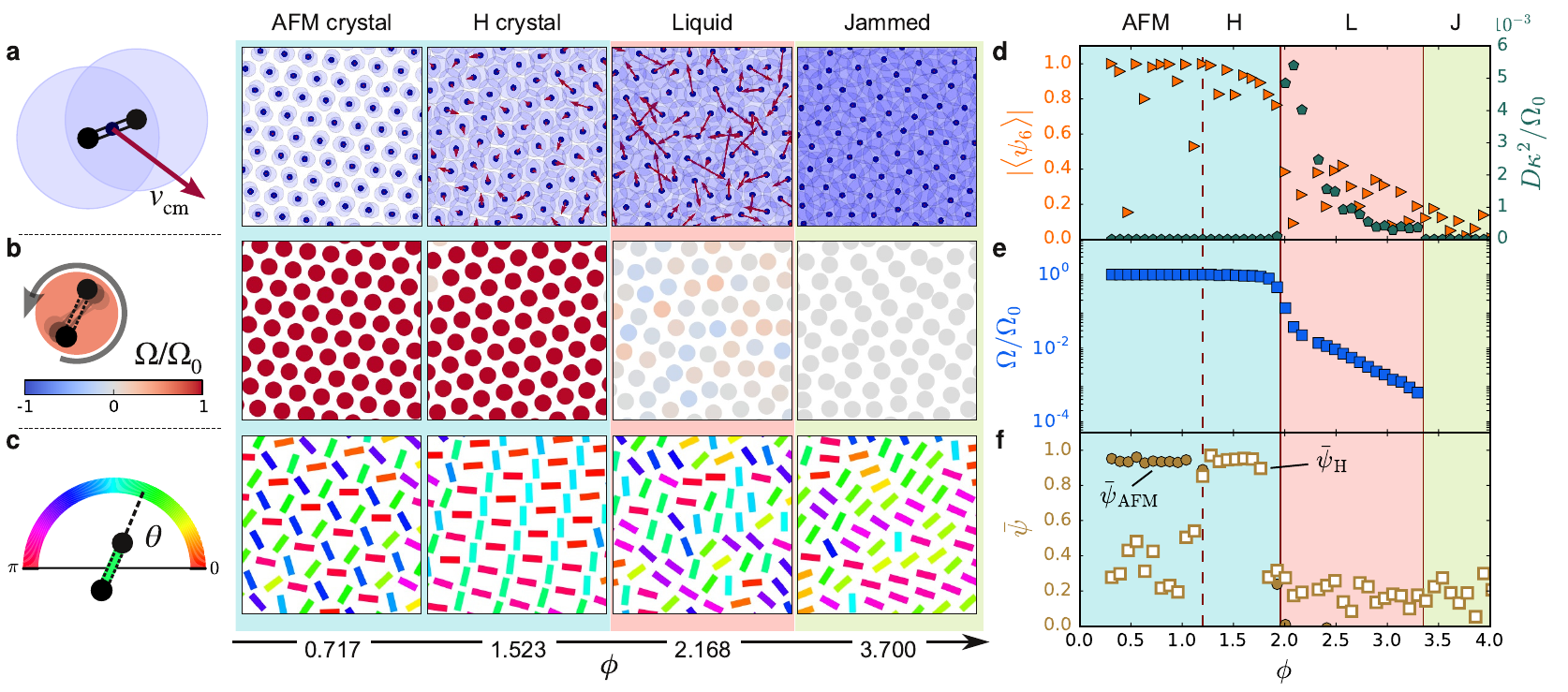}
  \caption{{\bfseries Bulk phases of the active spinner system.} The behaviour of
    dimer positions and orientations is investigated as a function of packing
    fraction $\phi$, for constant activity level $\alpha = 131.026$. Each row
    highlights different physical quantities of the system, shown schematically
    and displayed for simulation snapshots for four representative
    values of $\phi$ in {\bfseries a--c}. The snapshots cover roughly 10\% of the simulation area. 
    {\bfseries a}: Centre-of-mass position (dark dots)
    and velocities (red arrows), shown along with the soft exclusion area of
    individual charges (translucent discs);
    {\bfseries b}: angular rotation speed $\Omega/\Omega_0$;
    {\bfseries c}: orientation, represented by a
    fixed-length segment coloured by the angle made by the dimer with the $x$
    axis. Segment length does not represent actual
    dimer size. 
    {\bfseries d--f,} Ensemble measurements of steady-state physical quantities,
    as a function of $\phi$. 
    {\bfseries d}: bond-orientational order parameter, and diffusivity of dimer positions;
    {\bfseries e}: average angular speed. These quantities identify three distinct phases in
    different density ranges: crystal (blue background), liquid (red), and
    jammed (green). The rotational speed abruptly drops to zero (within numerical
    precision) in the jammed phase. 
    {\bfseries f}: order parameters quantifying Potts
    antiferromagnet ($\psiafm$) and striped herringbone
    ($\psih$) order in the phase relationships between rotating dimers
    in the crystal. 
  }
  \label{fig:phases}
\end{figure*}

Upon increasing the packing fraction beyond $\phimelt$, the diffusive and
spinning dynamics slow down as interactions become more prominent.
At $\phi = \phij \approx 3.3$, the diffusivity and spinning speed of the
ensemble both drop abruptly to zero, signifying a sharp transition from a liquid to a
frozen solid in which interactions completely overwhelm the external
drive~\cite{Fily2014}. As shown by representative snapshots (Fig.~2a--c,
fourth column) and the bond-orientational order parameter (Fig.~2d), the
dimer positions and orientations in the frozen state do not exhibit the ordering
of the crystalline phases. However, a different form of short-range
orientational order persists: dimers tend to form
ribbon-like assemblies which share a common alignment, see
Fig.~2c, fourth panel, and Supplementary Figure~S4. This
structure, which locally resembles smectic ordering in liquid crystals,
is a consequence of the constraints on tightly packing repulsive dimers.
The full description of this state, reminiscent of a
degenerate crystal~\cite{Frenkel1991}, goes beyond the scope of our work.

\section{Confinement-induced collective motion}
At a microscopic level, the bulk phases are distinguished by the relative
importance of rotational drive and orientation-dependent interactions. For a
steady state to be attained, torques must be balanced globally as well; in a
confined system, the overall torque may be balanced by viscous drag as well as
boundary forces. To investigate the interplay between rotational drive,
interactions, and confinement,
we simulated a system confined by a circular frictionless
boundary as depicted in Fig.~3a, for the same particle number
($N=768$), activity level and density range as in Fig.~2.
Densities are changed by varying the circle radius, since $\phi = NA/\pi R^2$.
Fig.~3a and Movie S1 show the dimer centre-of-mass
motion for three representative densities across different phases, all of which
display spontaneous macroscopic flows.

Measurements of the coarse-grained azimuthal velocity $v_\theta(r)$ as a
function of distance $r$ from the disc centre (see Methods) reveal qualitative
differences in the collective flows across phases. In both the crystal
($\phi=0.827$) and frozen ($\phi=3.750$) phases, the angular velocity about the
disc centre, $\omega(r)=v_\theta(r)/r$, is constant throughout the disc
(Fig.~3b), showing that the ensemble rotates around the centre
in unison as a rigid body. By contrast, the angular velocity profile is
nonuniform for the liquid ($\phi=2.395$), growing monotonically with distance
from the disc centre. These distinct behaviours persist over the entire phase diagram, as shown in
Fig.~3c which compares the steady-state values of
the flow angular velocity at the centre [$\omega(0)$] and edge [$\omega(R)$] of the disc as a
function of density. The centre and edge values coincide in the solid phases,
consistent with rigid-body rotation, whereas the liquid phase shows a persistent
enhancement of flow at the edge.
Collective vortical motion and boundary flows were previously demonstrated in
suspensions of swimming cells~\cite{Wioland2013,Lushi2014}. However, their spatial
structure and physical origin are profoundly different from the
confinement-induced flows reported here, which depend on the chiral activity of
the spinners as we now elucidate.

\begin{figure*}[t]
  \centering
  \includegraphics[width=4.3in]{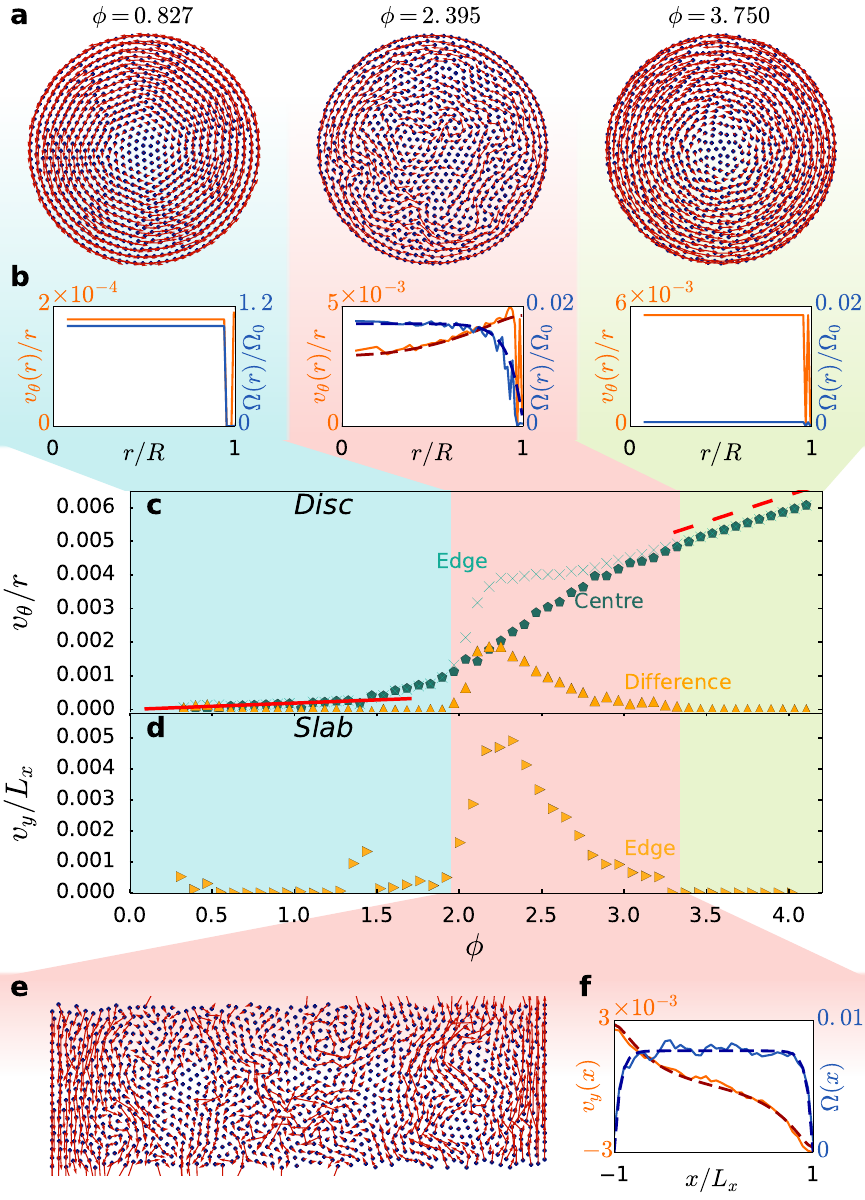}
  \caption{{\bf Collective motion reflects phase changes.}
    {\bf a,} Snapshots showing the drift of $N=768$ dimers
    confined by a circular boundary.
    Arrows indicate the displacements after $\Delta t=164/\Omega_0$ for
    $\phi=0.827$ (crystal), $\phi = 2.395$ (liquid), and
    $\phi=3.750$ (jammed). Arrows are scaled differently for visibility.
    {\bf b,} Time-averaged steady-state radial
    distributions of the orbital angular speed $\omega(r)$ about the disc centre (orange)
    and the local spin speed (blue), for the simulations shown in {\bf a}. Dashed
    lines are fits to the hydrodynamic theory.
    {\bf c,} Steady-state orbital angular speed $\omega(r)$ in simulation units as a function of
    density, measured at the disc centre ($r=0$) and edge ($r=R$). Coincidence
    of the two values is consistent with rigid-body rotation. The solid and
    dashed red lines show the theoretical prediction for the rigid-body rotation
    speed in the crystal and jammed phases respectively.
    {\bf d,} Steady-state tangential speed of dimers at the wall as a
    function of density, for $N=768$ dimers confined by two walls
    perpendicular to the $x$ direction and periodic boundary conditions along $y$.
    Density is varied by changing the area between the slabs while keeping the
    aspect ratio $L_y/L_x=2$ unchanged. 
    {\bf e,} Snapshot of dimer motion for 768 dimers confined between parallel
    slabs at $\phi=2.410$, with $L_y/L_x=1/3$.
    {\bf f,} Averaged steady-state velocity
    profile between the slabs (orange) and local spin speed (blue) for the
    simulation shown in {\bf e}. Dashed lines are fits to the hydrodynamic theory.
  } 
  \label{fig:current}
\end{figure*}

\subsection{Spontaneous collective rotation of rigid  phases}
The rigid-body rotation in the two solid phases, ordered and jammed, can be
understood by balancing torques about the centre of the circular boundary to
obtain an acceleration-free steady state. The
forces exerted by the boundary, being radially oriented, do not exert
torque. Thus, the driving torques acting on the dimers must be balanced by drag
forces. In the crystal interior, dimers homogeneously and steadily spin about
their individual centres at a rate $\Omega_0$, Fig.~3a, and the
resulting friction balances the driving torques at all times. However, the
spinning of the outermost layer of $N_\text{e}$ dimers is obstructed by the hard
boundary as shown schematically in Fig.~1e, which implies that the
driving torques on these dimers are not balanced 
by spinning. Rather, these torques drive an overall rotation of the crystal. The
corresponding  rigid-body rotation speed, $\omega_\text{rb}$, is
obtained by balancing the net drive $N_\text{e}\tau$ against the net drag torque
due to the rigid-body rotation which scales as $N \gamma \Omega R^2$, thereby
leading to $\omega_\text{rb} \sim (N_\text{e}/N) \tau/\gamma R^2 \propto \phi$.

In the frozen phase, local spinning of dimers relative to their neighbours is completely frustrated
by interactions. Therefore, the entire external torque $N\tau$ is balanced solely
by the drag due to orbital motion, giving rise to
$\omega_\text{rb} \sim \tau/\gamma R^2 \propto\phi$. The measured
rotation speeds quantitatively match the
predictions due to overall torque balance (solid and dashed lines in Fig.~3c).

\subsection{Emergent edge current in active spinner liquids}
The rigid-body motion of the two solid phases relies on the transmission of
torque \emph{via} shear stresses throughout the sample. If the disc is partitioned into
circular annuli, the net external drive acting on each annulus differs from the
net drag torque; neighboring annuli must exert shear forces on each other to
balance the total torque.
Unlike the  solid phases, the liquid cannot support a
shear stress through elastic deformations, which qualitatively explains the
absence of pure rigid-body rotation, Fig.~3a and b.
For a quantitative description of the emergent flow, we use a continuum theory
of an active chiral liquid coupled to a solid substrate. This phenomenological
model, introduced in Refs.~\onlinecite{Stark2005,Tsai2005}, generalizes the
so-called micropolar-fluid hydrodynamics~\cite{Scriven1961,Furthauer2012} by
including couplings to a frictional substrate.

Assuming incompressibility (as justified by the lack of significant spatial
variations in dimer density), the hydrodynamic description relies solely on
the conservation of momentum and angular momentum, and therefore involves two 
coarse-grained fields: the flow velocity
$\mathbf{v}(\mathbf{r})$, and the internal angular rotation, or spin, field
$\Omega(\mathbf{r})$. The hydrodynamic equations take on a compact form when
written in terms of $\Omega(\mathbf r)$ and the scalar vorticity, $\vort(\mathbf
r) = \frac{1}{2}{\hat {\mathbf z}}\cdot\nabla\times \mathbf {v(r)}$.
In the viscous steady-state limit these equations, which respectively amount
to local torque and force balance, are~\cite{Stark2005,Tsai2005}:
\begin{align}
  \label{eq:lubenskygollub1}
 D_\Omega \nabla^2\Omega - \Gamma^\Omega \Omega - \Gamma(\Omega-\vort)+\rho\tau &=0, \\
(4\eta+\Gamma) \nabla^2\vort -4\Gamma^v \vort -\Gamma \nabla^2 \Omega &= 0, 
  \label{eq:lubenskygollub2}
\end{align}
where $\rho$ is the active-spinner-fluid density, $\eta$ is the shear viscosity,
and $D_{\Omega}$ is a spin viscosity controlling the diffusive transport of
angular momentum. The coefficients $\Gamma^{\Omega}$ and $\Gamma^{v}$ quantify
the dissipation of angular and linear momentum respectively due to substrate
friction. The crucial spin-vorticity coupling is
embodied in the rolling friction $\Gamma$, which coarse-grains the frustration
between rotations and interactions outlined in Fig~1c.
Orientation-dependent interactions hinder the free spinning of adjacent fluid
elements, causing shear stresses proportional to $\Gamma$ unless the elements
flow past each other in such a way that the vorticity cancels the local spin.

Analysis of the hydrodynamic equations reveals that spatial variations
in the local spin field induce persistent flows. In the
absence of boundaries, the equations admit the flow-free solution $\Omega
=\rho\tau/(\Gamma^\Omega+\Gamma) = \Omega$, $\vort=0$. If a hard
boundary hinders spinning, however, $\Omega(\mathbf{r})$ varies from its value
imposed by the boundary to the constant interior value $\Omega$
over a length scale $\lambda_\Omega = \left [D_\Omega/(\Gamma
  +\Gamma_\Omega)\right]^{\frac{1}{2}}$ set by the competition between diffusion
and dissipation of local spin. The spatial variations in $\Omega$,
confined to the boundary, act as a source for vorticity which itself
decays over a length scale $\lambda_\vort = \left[(4\eta
  +\Gamma)/(4\Gamma^v)\right]^{\frac{1}{2}}$ set by drag.
These predictions match the simulation results, and a fit
to radially symmetric spin and flow fields (dashed lines in
Fig.~3b second panel) provides quantitative agreement with
four fitting parameters (more details in SI Text).

The spontaneous liquid flow only requires the obstruction of spinning by
the boundary, independently of its geometry. To highlight the robustness of
this emergent flow, we also study active spinner liquids in a slab
geometry with two edges aligned perpendicular to the $x$ axis and periodic
boundary conditions along $y$, as shown in Fig.~3e.  
This geometry suppresses rigid-body rotation in all phases; excess
driving torques are balanced by normal boundary forces.
Accordingly, no dimer motion is measured in the crystal and jammed phases,
Fig.~3d. However, a persistent flow parallel to the slab edges
arises in the liquid phase, demonstrating that the emergence of localized
shear flows at edges is a \emph{robust} feature of geometrically confined
active spinner liquids. The mechanism for the edge current
is the exchange between local spin and vorticity described above, which hinges
on the orientation dependence of dimer-dimer interactions. The hydrodynamic description
quantitatively reproduces the flow velocity profile $v_y(x)$ and spin field
$\Omega(x)$ (Fig.~3e, dashed lines).

\section{Conclusion}
Combining numerical simulations and analytical theory, we have elucidated the
phase behavior of interacting active spinners. The mutual frustration of
positional and time-periodic orientational order has been shown to yield a
variety of crystal and disordered phases.
Although we have focused on the density
dependence of the bulk and edge phenomena, the phases and their associated
emergent flows persist over a broad range of activity strengths (SI
Fig.~S5), which makes experimental realizations feasible.
Colloidal dumbbells~\cite{Manoharan2003,Sacanna2013} spun by 
phoretic stresses~\cite{Ebbens2016} or Quincke rotation~\cite{Bricard2013} would provide
a near-literal realization of our model. More broadly, the essential ingredients of
active spinners with orientation-dependent repulsive interactions are present in
a wide variety of experimental systems including chiral liquid crystals confined
to a monolayer and driven \emph{via} the Lehmann effect~\cite{Niton2013},
rotating nanorods propelled by biomolecular motors~\cite{Soong2000}, and
light-driven micromotors~\cite{Maggi2015}. We also envision macroscopic
realizations using chiral particles driven by airflow~\cite{Efrati2014} or
vibrations~\cite{Tsai2005}, with soft
interactions provided by electrostatic or magnetic repulsion. Besides opening up
new avenues to explore nonequilibrium physics in simple settings, the novel
phases arising from the interplay between interactions and
spinning may be exploited for tunable torque transmission~\cite{Williams2015} or
for self-assembly of anisotropic particles into ordered patterns. 

{\section{Materials and Methods}
  Details of the molecular dynamics simulations including implementation of
  dimers and boundaries, ensemble averaging and
  spatial coarse-graining of relevant physical quantities, and order
  parameters used to distinguish various phases are provided in SI Text.
}

{\bfseries Acknowledgments: }{We thank Bryan G. Chen and Thomas H. Beuman for useful discussions.
  B.C.v.Z., J.P., and V.V. were funded by FOM, a VIDI grant from NWO, and a
  Delta ITP Zwaartekracht grant. W.T.M.I. acknowledges support from the Packard
  foundation through a Packard Fellowship, and from the NSF CAREER Program
  (DMR-1351506). D.B. acknowledges support from ANR grant MiTra.}

\clearpage
\setcounter{figure}{0}
\setcounter{equation}{0}
\setcounter{section}{0}

{\sffamily \Large {\bfseries Supplementary Information}}

\renewcommand{\theequation}{S\arabic{equation}}
\renewcommand{\thefigure}{S\arabic{figure}}

\section{Materials and methods}
\subsection*{Simulations} Our molecular dynamics simulations solve Newton's
equations for a system of point particles with specified pairwise interactions,
external forces, and drag coefficients. Particles interact with a pairwise
repulsive Yukawa potential with identical charge $b$ and screening length
$\kappa$, enabled for all particle pairs with separation $r < 10\kappa^{-1}$.
Dimers are created by connecting pairs of particles with stiff harmonic springs
of equilibrium length $d$ and spring constant $k = 10^4 b\kappa^3$. Torques are
applied \emph{via} an external force $F=\tau/d$ oriented perpendicular to the
link at all times. Each point particle also experiences a drag force
proportional to velocity with coefficient $-\gamma$. Simulations are initialized
with dimers at random positions and orientations within the simulation box.
Particle positions and velocities are updated by integrating Newton's equations
using a symplectic Euler method with time step $\Delta t = 0.0086/\Omega_0$. A
typical simulation runs for $10^7$ time steps, taking roughly 100 CPU hours for
system size $N=768$ at the highest densities, with a snapshot of dimer data
saved every $10^3$ steps. Ensemble averages are carried out over the final 8000
snapshots.

Confining boundaries are implemented using a steep one-sided harmonic repulsive
potential $V(x) = k_\text{w}x^2/2$ experienced by all particles, where $x$ is
the penetration distance into the boundary, and $k_\text{w} = 3.14\times
10^2b\kappa^3$. For simulations confined by a circular boundary, coarse-grained
fields of the form $f(r)$ are computed by dividing the simulation region into 39
concentric annuli with widths inversely proportional to their mean radius $r$ so
that the number of dimers is the same in each annulus on average. The relevant
quantity averaged over all dimers occupying the annulus at $r$ provides a
discretized numerical estimate of the coarse-grained field value $f_p(r)$ in
frame $p$. The estimate is then averaged over the final 8000 snapshots to obtain
the coarse-grained field $f(r) = \langle f_p(r) \rangle_p$. A similar averaging
provides the coarse-grained fields $\Omega(x)$ and $v_y(x)$ for the slab
geometry, but with the simulation area between the slabs divided into 40 strips
with edges parallel to the $y$ axis. Averaging the relevant quantity over dimers
occupying a strip centred at $x$ provides the discrete coarse-grained field
value $f_p(x)$.

\subsection*{Order parameters}
The local bond-orientational order parameter ${\psi_6}_{,i} = \sum_{j=1}^{n_i}
e^{6\mathrm{i}\theta_{ij}}/n_i$, where $j$ indexes the $n_i$ nearest neighbours
of $i$ and $\theta_{ij}$ is the angle made by the bond connecting $i$ and $j$
with the $x$ axis, measures the extent to which the neighbours of dimer $i$
match the orientational order of the triangular lattice. The global order
parameter $|\langle \psi_6 \rangle| = |\sum_{i=1}^N {\psi_6}_{,i} / N|$ measures
the extent to which local bond orientations are aligned across the system. A
perfect triangular lattice has $|\langle \psi_6 \rangle|=1$.
  
The local order parameters ${\psiafm}_{,i}$ and ${\psih}_{,i}$ report whether
the orientations of dimer $i$ and its nearest neighbours $j$ (identified
\emph{via} a Delaunay triangulation) are consistent with the expected phase
differences for the 3P-AFM and H crystal phases respectively. To identify the
3P-AFM phase, we check whether orientation differences between neighbours are
$\pm \pi/3$, by computing
  \begin{equation}
    \label{eq:psiafm} {\psiafm}_{,i} = \frac{1}{z_i} \sum_{j=1}^{z_i}
1-\frac{1}{3}\left[4 \cos^2(\theta_{ij})-1\right]^2,
  \end{equation} where $\theta_{ij} = \theta_i-\theta_j$ and $z_i$ is the number
of neighbours of dimer $i$. The expression evaluates to 1 if
$[(\theta_i-\theta_j) \mod \pi] \in \{\pi/3,2\pi/3\}$ for all neighbours, and
has an expectation value of 0 if angle differences are randomly distributed.

  For the H phase, we first arrange the neighbours in order of increasing angle
made by the link connecting $i$ and $j$ with the $x$ axis. Our goal is to
evaluate the closeness of all possible circular shifts of this neighbour
arrangement with the sequence $S\equiv\{0, \pi/2, \pi/2, 0, \pi/2, \pi/2\}$. We
define the shift $k$ as the integer in $\{0,1,2\}$ which minimizes
$\sin^2(\theta_{ik}) + \sin^2(\theta_i-\theta_{k+3})$ in the ordered
arrangement. The local order parameter is then computed \emph{via}
  \begin{equation}
    \label{eq:psih} {\psih}_{,i} = -\frac{1}{z_i} \sum_{j=1}^{z_i} \cos \left(
      \begin{cases} 2\theta_{ij} +\pi, & \text{if } j\mod 3 = k \\ 2\theta_{ij},
& \text{otherwise}
      \end{cases} \right),
  \end{equation} which evaluates to 1 only if the sequence of $\theta_{ij}$
starting from $j = k$ matches $S$ and is close to zero for a random distribution
of dimer orientations.

  Under periodic boundary conditions, the crystals form phase-locked domains
separated by defects and grain boundaries which bring down the value of the
order parameters from 1 when averaged over all points. In Fig.~2f, we identify
the predominant local order within domains by plotting the most probable values
$\bar{\psi}_\text{AFM}$ and $\bar{\psi}_\text{H}$. These are obtained by binning
the local values $\psi_i$ from every 50th frame in the range $8000\leq p \leq
10000$ into 20 equally spaced bins, and reporting the coordinate of the bin with
highest occupancy.

\section{Effective interaction between dimer pairs}
In the limit that the dimer length $d$ is small compared to the dimer
separation, each dimer can be considered a superposition of a monopole carrying
the net charge $2b$, and a quadrupole charge distribution. Therefore, the
interaction between a pair of dimers can be written as a sum of
monopole-monopole, monopole-quadrupole, and quadrupole-quadrupole terms. The
monopole-monopole contribution is independent of dimer orientation. Suppose
the angle made by dimer $i$ evolves in time as $\theta_i=\Omega_0t+\delta_i$. By
symmetry considerations, the monopole-quadrupole contribution integrates to a
quantity which is independent of the phases $\delta_i$. The quadrupole-quadrupole
contribution does depend on the relative phases, and has the form
\begin{equation}
  \label{eq:quadquad}
  E_{ij}=J(r_{ij})\cos(2\theta_i-2\theta_j)+K(r_{ij})\cos(2\theta_i+2\theta_j-4\phi_{ij}),
\end{equation}
where $\phi_{ij}$ is the angle made by the link connecting $i$ and $j$ with the
$x$-axis, and $J,K$ are functions of the centre-of-mass separation $r_{ij}$, set
by the Yukawa parameters:
\begin{widetext}
\begin{align}
  J(r) &=  \frac{b}{128 r} \left(\frac{d}{r}\right)^4 e^{-\kappa r}
  [9(1+\kappa r) +5 (\kappa r)^2 + 2(\kappa r)^3 + (\kappa r)^4], \\
  K(r) &= \frac{b}{128 r} \left(\frac{d}{r}\right)^4 e^{-\kappa r}
  [105(1+\kappa r) +45 (\kappa r)^2 + 10(\kappa r)^3 + (\kappa r)^4]
\end{align}  
\end{widetext}
  
For the rotating dipoles with constant angular speed
$\Omega_0$ with fixed centre-of-mass positions separated by the lattice spacing $a$, we have
\begin{equation}
  \label{eq:quadrot}
  E_{ij}=J(a)\cos(2\delta_i-2\delta_j)+K(a)\cos(4\Omega_0t + 2\delta_i+2\delta_j-4\phi_{ij}).
\end{equation}
When the energy is integrated over a cycle, the second term integrates to zero,
and hence the average potential energy over the cycle is  $(1/T)\int_0^T E_{ij}\, dt
= J(a)\cos(2\delta_i-2\delta_j)$. 

\section{Hydrodynamic model: rescaling and approximate solution}
In this section, we derive closed-form approximate solutions to the hydrodynamic
equations, Eqs.~4--5 of the main text, which are useful for numerical fitting to
the spin and velocity profiles of the active spinner liquid under confinement.
We follow Ref.~\onlinecite{Tsai2005} and
introduce lengths via $\lambda_\Omega^2= D_\Omega/(\Gamma +\Gamma_\Omega)$ and
$\lambda_\vort^{-2} = 4\Gamma^v/(4\eta +\Gamma)$, and unitless parameters $p =
\Gamma/(\Gamma +\Gamma^\Omega)$ and $q = \Gamma/(4\eta +\Gamma)$. Then the
equations become 
\begin{align}
  \label{eq:lubenskygollubrescale}
 (\lambda_\Omega^2\nabla^2-1)\Omega + p\vort +\tilde{\tau} &=0, \\
 (\nabla^2-\lambda_\vort^{-2})\vort - q \nabla^2 \Omega &= 0,  
\end{align}
where $\tilde{\tau} = \rho\tau/(\Gamma^\Omega+\Gamma) \sim \tau/(\gamma_\Omega +
\Gamma/\rho)$. In the interior of a sample, away from the edges, we expect (and observe)
$\Omega \approx \tilde{\tau}-p\vort$.  We also observe numerically that
$p\vort$ is negligible compared to the other two terms. Since $\langle{\Omega}
\rangle \ll \Omega_0$ in the liquid phase, this implies $\Gamma \gg
\Gamma^\Omega \Rightarrow p \approx 1$. With these simplifications, and the
requirement of zero spin and zero tangential forces at the boundary,
we get a closed-form solution for the two hydrodynamic fields. In the slab geometry,
with slab boundaries at $x=\pm L/2$, they have the form:
\begin{align}
\Omega(x) &= \tilde\tau  \left[1-\text{sech}\left(\frac{L}{2 \lambda _{\Omega
      }}\right) \cosh \left(\frac{x}{\lambda _{\Omega }}\right)\right], \\
\vort(x) &= \frac{q \tilde\tau 
  \left[\text{sech}\left(\frac{L}{2 \lambda_{\vort }}\right)
    \cosh \left(\frac{x}{\lambda _{\vort}}\right)-\text{sech}\left(\frac{L}{2 \lambda _{\Omega }}\right) \cosh\left(\frac{x}{\lambda _{\Omega }}\right)\right]}{1-\lambda _{\Omega }^2/\lambda _{\vort }^2}.
\end{align}
The corresponding velocity field is obtained by integrating the vorticity. The
current magnitude at the edges ($x=\pm L/2$) is
\begin{equation}
  \label{eq:edgeslabcurrent}
  v_\text{edge} = \frac{2 q \tilde\tau  \left[\lambda _{\vort }
      \tanh
      \left(\frac{L}{2 \lambda _{\vort }}\right)-\lambda _{\Omega } \tanh
      \left(\frac{L}{2 \lambda _{\Omega }}\right)\right]}{1-\lambda _{\Omega }^2/\lambda _{\vort }^2 }.
\end{equation}

Using these results, we can extract the values of the lengths and dimensionless
parameters from the simulations. We first fit the spin field $\Omega(x)$ since
the decay length $\lambda_\Omega$ tends to be much smaller than
$\lambda_\vort$, allowing the former to be fit accurately for narrow slabs
where the width might be comparable to the latter. The fit to the spin
field fixes the parameters $\tilde\tau$ and $\lambda_\Omega$. The second fitting
of the velocity field then fixes the remaining two parameters $q$ and
$\lambda_\vort$.

The parameter values obtained from the fit for the slab simulation in main text
Fig.~3f, with $L=121.43 d$, are $\tilde\tau=0.06627, q=0.0150,
\lambda_\Omega=3.554 d, \lambda_\vort = 22.65 d$. The approximation that $p\vort
\ll \Omega$ requires $q \ll 1$, satisfied by the fit.

The same procedure is used for a liquid confined to a disc of radius $R$, for
which the approximate radially symmetric solution for the spin and vorticity
fields is
\begin{widetext}
\begin{align}
  \Omega(r) &=\tilde\tau \left[1- \frac{R \left(1-
        \lambda_{\Omega}^2/\lambda_{\vort}^2\right)
      I_2\left(\frac{R}{\lambda_{\vort}}\right)
      I_0\left(\frac{r}{\lambda_{\Omega} }\right)}{2 b \lambda_{\vort}
      I_1\left(\frac{R}{\lambda_{\vort}}\right)
      I_2\left(\frac{R}{\lambda_{\Omega}
        }\right)+I_2\left(\frac{R}{\lambda_{\vort}}\right) \left(R
        (1-\lambda_{\Omega}^2/\lambda_{\vort}^2 )
        I_0\left(\frac{R}{\lambda_{\Omega} }\right)-2 b \lambda_{\Omega}
        I_1\left(\frac{R}{\lambda_{\Omega} }\right)\right)} \right], \\
  \vort(r) &= \frac{b  R \tilde\tau
    \left[I_0\left(\frac{r}{\lambda_{\vort} }\right)
      I_2\left(\frac{R}{\lambda_{\Omega}
        }\right)-I_0\left(\frac{r}{\lambda_{\Omega} }\right)
      I_2\left(\frac{R}{\lambda_{\vort} }\right)\right]}{2 \lambda_{\vort}
    \left((b-1)+\lambda_{\Omega} ^2/\lambda_{\vort} ^2\right)
    I_0\left(\frac{R}{\lambda_{\Omega} }\right)
    I_1\left(\frac{R}{\lambda_{\vort} }\right)+I_0\left(\frac{R}{\lambda_{\vort}
      }\right) \left[R (1-\lambda_{\Omega}^2/\lambda_{\vort}^2 ) I_0\left(\frac{R}{\lambda_{\Omega} }\right)-2 b
      \lambda_{\Omega}  I_1\left(\frac{R}{\lambda_{\Omega}
        }\right)\right]}.
\end{align}
\end{widetext}
Here, $I_m$ is the modified Bessel function of first kind of order $m$.
The parameters obtained from the fit to the approximate solution for the
disc simulation in Fig.~3b of the main text ($\phi = 2.395,\,R = 33.65 d$) are 
 $\tilde\tau = 0.1475, q=0.0161, \lambda_\Omega = 2.484 d, \lambda_\vort=15.61 d$.

\begin{figure}
  \centering
  \includegraphics[width=\columnwidth]{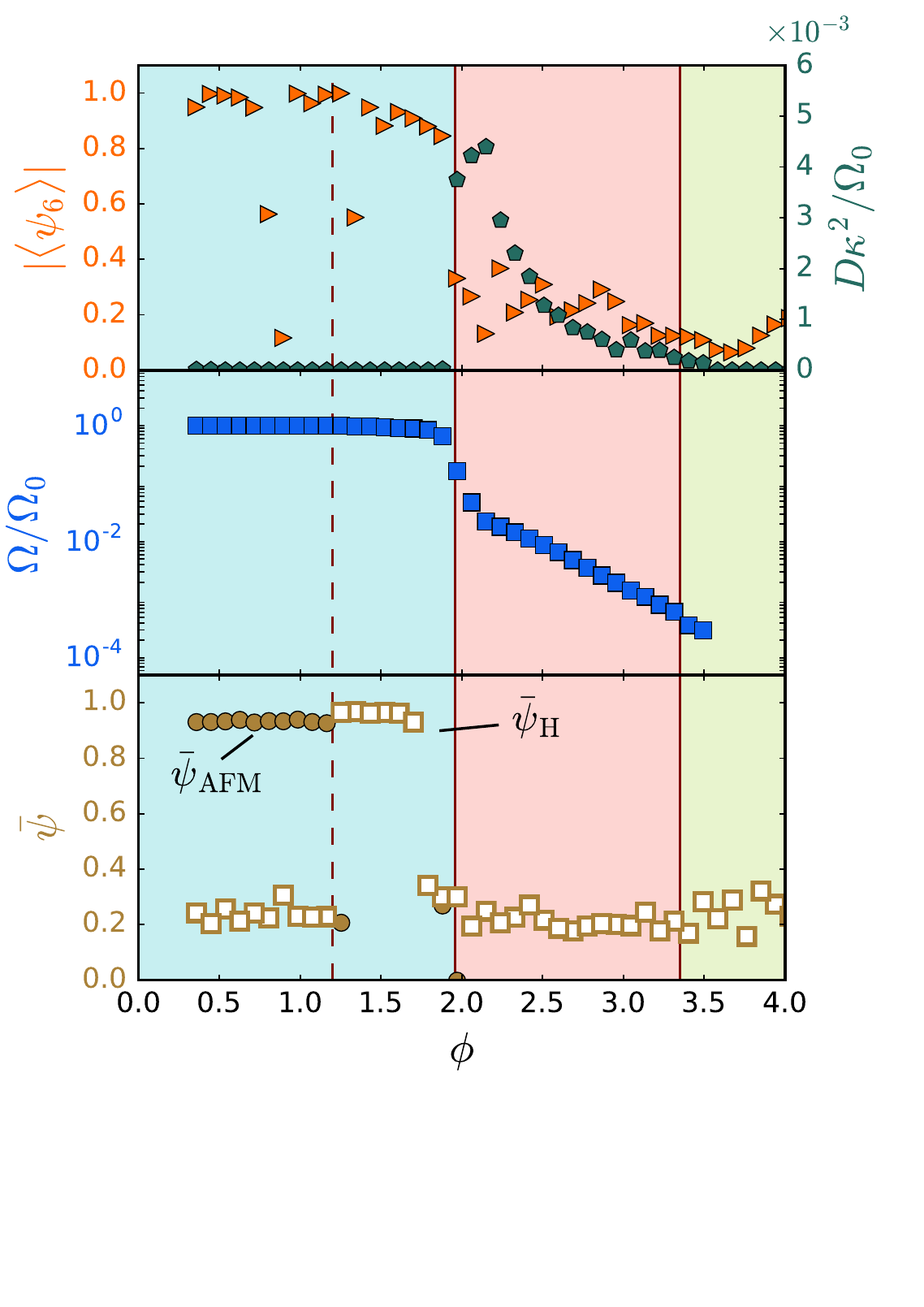}
  \caption{Ensemble measurements of steady-state physical quantities
    as a function of $\phi$ for a system with $N=3072$ dimers, four times the
    system size of the simulations reported in Fig. 2 of the main text. 
    {\bfseries Top row}: bond-orientational order parameter, and diffusivity of dimer positions;
    {\bfseries Middle row}: average angular speed. These quantities identify three distinct phases in
    different density ranges: crystal (blue background), liquid (red), and
    jammed (green). The rotational speed abruptly drops to zero (within numerical
    precision) in the jammed phase. 
    {\bfseries Bottom row}: order parameters quantifying Potts
    antiferromagnet ($\psiafm$) and striped herringbone
    ($\psih$) order in the phase relationships between rotating dimers
    in the crystal.
    The vertical lines are at the same values of $\phi$ as in Fig. 2 of the main
    text. The density ranges for the distinct phases are almost identical for
    $N=768$ and $N=3072$. The transition from liquid to jammed occurs at a
    slightly higher density, $\phij \approx 3.5$ for $N=3072$ (compared to
    $\phij \approx 3.3$ for $N=768$). 
  }
  \label{fig:3072}
\end{figure}

\begin{figure}
  \centering
  \includegraphics[width=\columnwidth]{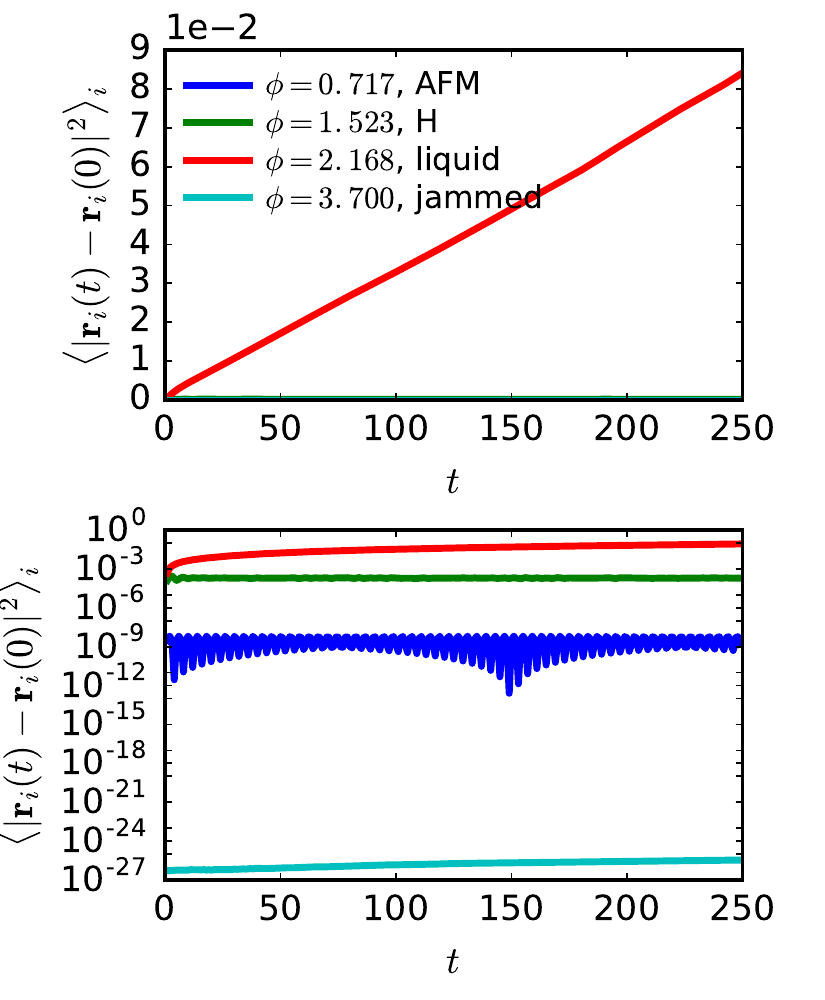}
  \caption{Mean squared displacement of dimers over time, for the periodic
    system under four representative densities which are the same as in Fig. 2
    of the main text. The MSD is calculated independently for 76 randomly chosen
    dimers (one-tenth of the total) and averaged over them. The bottom plot
    shows the same data on a logarithmic scale. The
    system in the liquid phase ($\phi=2.168$) has diffusive dimer dynamics with
    the mean square displacement growing linearly with time. Dimers in the crystal
    ($\phi=0.717$ and $\phi=1.523$) and jammed ($\phi=3.700$) states do not
    diffuse over long times. However, in the crystal phases
    there is a finite displacement at short times reflecting the vibrations of
    the dimers around their mean positions which are fixed over time. These
    vibrations are  larger in the H phase compared to the AFM phase.}
  \label{fig:msd}
\end{figure}

\begin{figure}
  \centering
  \includegraphics[width=\columnwidth]{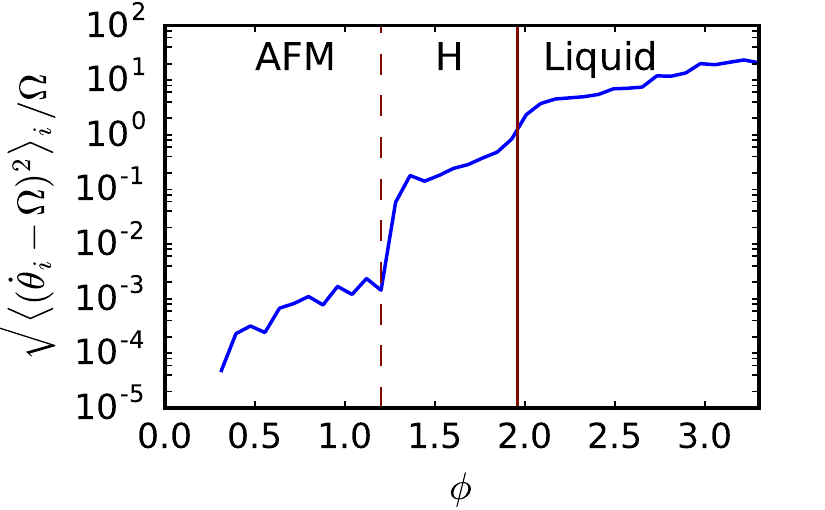}
  \caption{Fluctuations in dimer spin velocities $\dot{\theta}_i$ around their
    mean value $\Omega$ in the steady state for the simulations under periodic
    boundary conditions with $\alpha = 131,\,\beta=0.0075$. The fluctuations are
    quantified by the standard deviation of spin velocities, normalized by their
    mean. The normalized fluctuations are negligible in the 3P-AFM crystal phase
    ($\phi < 1.2$) and small in the H crystal phase ($1.2 < \phi < 1.9$). They
    become very large in the liquid phase ($\phi > 1.9$), showing that dimers no
    longer rotate uniformly in the liquid; their rotational dynamics are
    dominated by interactions, which change constantly as dimers diffuse through
    the liquid. In the jammed phase ($\phi > 3.3$), the mean spin velocity
    $\Omega = 0$ and the normalized spread in spin velocities is undefined. }
  \label{fig:spinfluct}
\end{figure}

\begin{figure}
  \centering
  \includegraphics{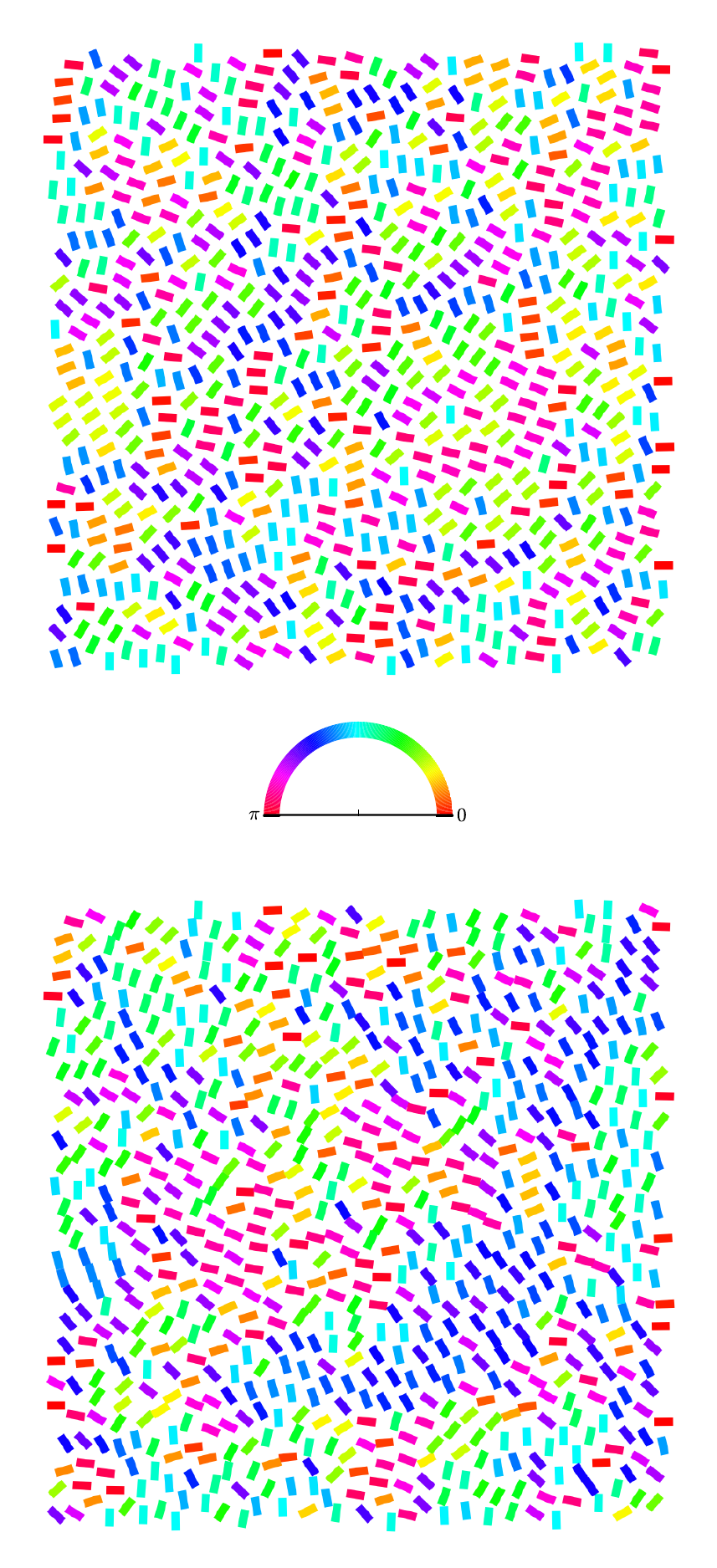}
  \caption{Snapshot of a simulation in the jammed phase ($\alpha = 131.026, \phi=3.7$) coloured by (left) orientation angle and (right) orientation angle multiplied by 3. Regions of uniform colour on the right indicate regions where dimers are aligned modulo $\pi/3$.}
  \label{fig:si_glass_order}
\end{figure}

\begin{figure}
 \centering
 \includegraphics{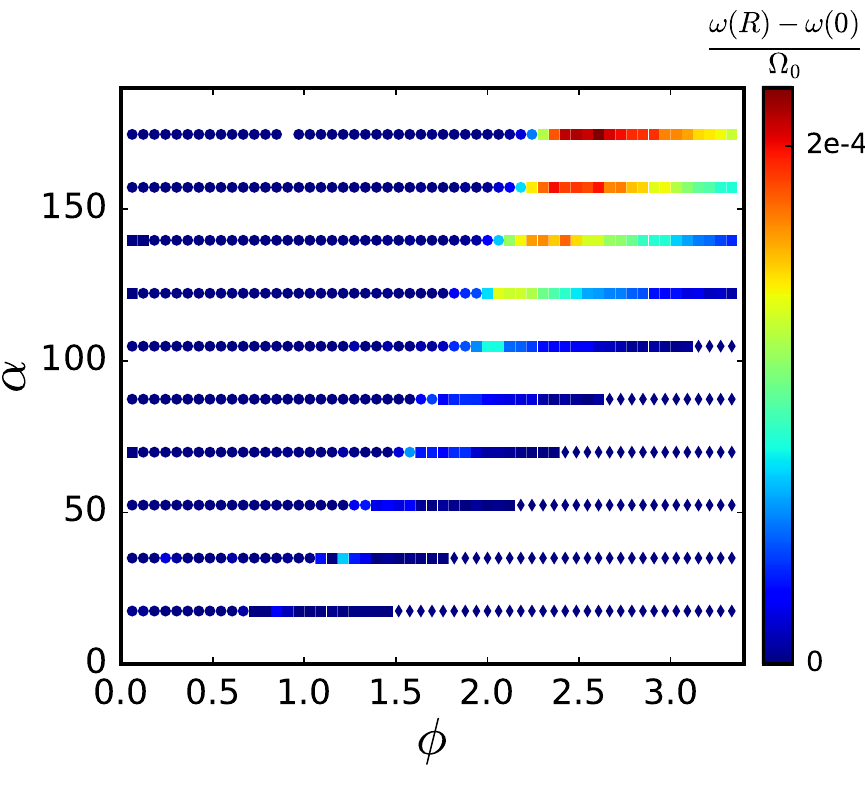}
 \caption{
   {\bfseries Bulk phases are reflected in edge currents and persist over a
     range of torques.}
   The flow at the edge plotted as a function of activity level $\alpha$
   and packing fraction $\phi$ in the disc geometry. In these simulations, the
   disc radius $R=48.27 d$ was kept constant and the density was varied by
   changing the number of dimers confined to the disc. The tangential flow
   velocity $v_\theta(r)$ was measured as for the systems in main text Fig.~3,
   and the edge flow is reported as the difference in $\omega(r) =
   v_\theta(r)/r$ at the edge and centre of the disc.
   Symbols are coloured by edge current magnitude, whereas symbol shapes
   are chosen by values of bulk properties that identify the different phases.
   {\bfseries Circles,} Active rotator crystals, identified by $\langle \Omega
   \rangle > 0.4$ and $\langle \psi_6 \rangle > 0.45$.
   {\bfseries Diamonds,} Frozen phase, identified by $\langle \Omega \rangle <
   10^{-5}$.
   {\bfseries Squares,} Dynamic and disordered liquid-like states, that do not
   satisfy the previous criteria for the crystal and frozen phases.
 }
 \label{fig:phasediagram}
\end{figure}


\begin{thebibliography}{10}

\bibitem{Marchetti_review}
Marchetti MC et~al. (2013) {Hydrodynamics of soft active matter}.
\newblock {\em Rev. Mod. Phys.} 85(3):1143--1189.

\bibitem{Cavagna_review}
Cavagna A, Giardina I (2014) {Bird Flocks as Condensed Matter}.
\newblock {\em Ann. Rev. Cond. Matt. Phys.} 5(1):183--207.

\bibitem{Vicsek_review}
Vicsek T, Zafeiris A (2012) {Collective motion}.
\newblock {\em Phys. Rep.} 517(3-4):71--140.

\bibitem{Dogic2013}
Sanchez T, Chen DTN, DeCamp SJ, Heymann M, Dogic Z (2013) {Spontaneous motion
  in hierarchically assembled active matter}.
\newblock {\em Nature} 491(7424):431--434.

\bibitem{Bausch2010}
Schaller V, Weber C, Semmrich C, Frey E, Bausch AR (2010) {Polar patterns of
  driven filaments}.
\newblock {\em Nature} 467(7311):73--77.

\bibitem{Bricard2013}
Bricard A, Caussin JB, Desreumaux N, Dauchot O, Bartolo D (2013) {Emergence of
  macroscopic directed motion in populations of motile colloids.}
\newblock {\em Nature} 503(7474):95--8.

\bibitem{Palacci2013}
Palacci J, Sacanna S, Steinberg AP, Pine DJ, Chaikin PM (2013) {Living crystals
  of light-activated colloidal surfers}.
\newblock {\em Science} 339(6122):936--940.

\bibitem{Ginot2014}
Ginot F et~al. (2015) Nonequilibrium equation of state in suspensions of active
  colloids.
\newblock {\em Phys. Rev. X} 5:011004.

\bibitem{Thutupalli2011}
Thutupalli S, Seemann R, Herminghaus S (2011) {Swarming behavior of simple
  model squirmers}.
\newblock {\em New J. Physics} 13:073021.

\bibitem{Deseigne2010}
Deseigne J, Dauchot O, Chat\'e H (2010) Collective motion of vibrated polar
  disks.
\newblock {\em Phys. Rev. Lett.} 105:098001.

\bibitem{Kudrolli2010}
Kudrolli A (2010) Concentration dependent diffusion of self-propelled rods.
\newblock {\em Phys. Rev. Lett.} 104:088001.

\bibitem{TonerTu}
Toner J, Tu Y (1995) {Long-range order in a two-dimensional dynamical XY model:
  how birds fly together}.
\newblock {\em Phys. Rev. Lett.} 75(23):4326--4329.

\bibitem{Cates_review}
Tailleur J, Cates ME (2015) {Motility-Induced Phase Separation}.
\newblock {\em Ann. Rev. Cond. Matt. Phys.} 6:219--244.

\bibitem{Toner_review}
Toner J, Tu Y, Ramaswamy S (2005) {Hydrodynamics and phases of flocks}.
\newblock {\em Annals of Physics} 318(1):170--244.

\bibitem{Snezhko2016}
Snezhko A (2016) {Complex collective dynamics of active torque-driven colloids
  at interfaces}.
\newblock {\em Current Opinion in Colloid and Interface Science} 21:65--75.

\bibitem{Drescher2009}
Drescher K et~al. (2009) {Dancing Volvox : Hydrodynamic Bound States of
  Swimming Algae}.
\newblock {\em Physical Review Letters} 102(16):168101.

\bibitem{Petroff2015}
Petroff AP, Wu Xl, Libchaber A (2015) {Fast-Moving Bacteria Self-Organize into
  Active Two-Dimensional Crystals of Rotating Cells}.
\newblock {\em Phys. Rev. Lett.} 114:158102.

\bibitem{Denk2016}
Denk J, Huber L, Reithmann E, Frey E (2016) {Active Curved Polymers Form Vortex
  Patterns on Membranes}.
\newblock {\em Phys. Rev. Lett.} 116(17):178301.

\bibitem{Riedel2005}
Riedel I, Kruse K, Howard J (2005) {A self-organized vortex array of
  hydrodynamically entrained sperm cells}.
\newblock {\em Science} 309(5732):300.

\bibitem{Sumino2102}
Sumino Y et~al. (2012) {Large-scale vortex lattice emerging from collectively
  moving microtubules}.
\newblock {\em Nature} 483(7390):448--452.

\bibitem{Tsai2005}
Tsai JC, Ye F, Rodriguez J, Gollub JP, Lubensky TC (2005) {A Chiral Granular
  Gas}.
\newblock {\em Phys. Rev. Lett.} 94(21):214301.

\bibitem{Saglimbeni2015}
Saglimbeni F, Dipalo M, De~Angelis F, Di~Leonardo R, Maggi C (2015)
  {Micromotors with asymmetric shape that efficiently convert light into work
  by thermocapillary effects}.
\newblock {\em Nat Commun} 6:1--5.

\bibitem{Tabe2003}
Tabe Y, Yokoyama H (2003) {Coherent collective precession of molecular rotors
  with chiral propellers}.
\newblock {\em Nature Materials} 2(12):806--809.

\bibitem{Oswald2015}
Oswald P, Poy G (2015) {Lehmann rotation of cholesteric droplets: Role of the
  sample thickness and of the concentration of chiral molecules}.
\newblock {\em Phys. Rev. E} 91(3):032502.

\bibitem{Lemaire2008}
Lemaire E, Lobry L, Pannacci N, Peters F (2008) {Viscosity of an
  electro-rheological suspension with internal rotations}.
\newblock {\em Journal of Rheology}.

\bibitem{Furthauer2013a}
F{\"{u}}rthauer S, Strempel M, Grill SW, J{\"{u}}licher F (2013) {Active Chiral
  Processes in Thin Films}.
\newblock {\em Phys. Rev. Lett.} 110(4):048103.

\bibitem{Uchida2010}
Uchida N, Golestanian R (2010) Synchronization and collective dynamics in a
  carpet of microfluidic rotors.
\newblock {\em Phys. Rev. Lett.} 104:178103.

\bibitem{Uchida2010a}
Uchida N, Golestanian R (2010) {Synchronization in a carpet of hydrodynamically
  coupled rotors with random intrinsic frequency}.
\newblock {\em Europhys Lett} 89(5):50011.

\bibitem{Acebron2005}
Acebr\'on JA, Bonilla LL, P\'erez~Vicente CJ, Ritort F, Spigler R (2005) The
  kuramoto model: A simple paradigm for synchronization phenomena.
\newblock {\em Rev. Mod. Phys.} 77:137--185.

\bibitem{Nonaka2002}
Nonaka S, Shiratori H, Saijoh Y, Hamada H (2002) {Determination of left–right
  patterning of the mouse embryo by artificial nodal flow}.
\newblock {\em Nature} 418(6893):96--99.

\bibitem{Guirao2010}
Guirao B et~al. (2010) {Coupling between hydrodynamic forces and planar cell
  polarity orients mammalian motile cilia}.
\newblock {\em Nature Cell Biology} 12(4):341--350.

\bibitem{Button2012}
Button B et~al. (2012) A periciliary brush promotes the lung health by
  separating the mucus layer from airway epithelia.
\newblock {\em Science} 337(6097):937--941.

\bibitem{Brumley2015}
Brumley DR, Polin M, Pedley TJ, Goldstein RE (2015) {Metachronal waves in the
  flagellar beating of Volvox and their hydrodynamic origin}.
\newblock {\em Journal of The Royal Society Interface} 12(108):20141358.

\bibitem{Kirchhoff2005}
Kirchhoff R, L{\"{o}}wen H (2005) {T-structured fluid and jamming in driven
  Brownian rotators}.
\newblock {\em Europhys Lett} 69(2):291--297.

\bibitem{Kaiser2013}
Kaiser A, L{\"{o}}wen H (2013) {Vortex arrays as emergent collective phenomena
  for circle swimmers}.
\newblock {\em Physical Review E} 87(3):032712.

\bibitem{Lenz2003}
Lenz P, Joanny JF, J\"ulicher F, Prost J (2003) Membranes with rotating motors.
\newblock {\em Phys. Rev. Lett.} 91:108104.

\bibitem{Nguyen2014}
Nguyen NHP, Klotsa D, Engel M, Glotzer SC (2014) {Emergent Collective Phenomena
  in a Mixture of Hard Shapes through Active Rotation}.
\newblock {\em Phys. Rev. Lett.} 112(7):075701.

\bibitem{Sabrina2015}
Sabrina S, Spellings M, Glotzer SC, Bishop KJM (2015) {Coarsening dynamics of
  binary liquids with active rotation}.
\newblock pp. 1--9.

\bibitem{Spellings2015}
Spellings M et~al. (2015) {Shape control and compartmentalization in active
  colloidal cells}.
\newblock {\em Proc. Natl. Acad. Sci. USA} 112(34):E4642--E4650.

\bibitem{Yeo2015}
Yeo K, Lushi E, Vlahovska PM (2015) Collective dynamics in a binary mixture of
  hydrodynamically coupled microrotors.
\newblock {\em Phys. Rev. Lett.} 114:188301.

\bibitem{Goto2015}
Goto Y, Tanaka H (2015) {Purely hydrodynamic ordering of rotating disks at a
  finite Reynolds number}.
\newblock {\em Nat Commun} 6:5994.

\bibitem{Aragones2016}
Aragones JL, Steimel JP, Alexander-Katz A (2016) {Elasticity-induced force
  reversal between active spinning particles in dense passive media.}
\newblock {\em Nat Commun} 7:11325.

\bibitem{coq2011}
Coq N et~al. (2011) Collective beating of artificial microcilia.
\newblock {\em Phys. Rev. Lett.} 107:014501.

\bibitem{Yan2015}
Yan J, Bae SC, Granick S (2015) {Rotating crystals of magnetic Janus colloids}.
\newblock {\em Soft Matter} 11(1):147--153.

\bibitem{Schick1977}
Schick M, Griffiths RB (1977) {Antiferromagnetic ordering in the three-state
  Potts model}.
\newblock {\em J. Phys. A: Math Gen} 10(12):2123--2131.

\bibitem{Mouritsen1982}
Mouritsen OG, Berlinsky AJ (1982) {Fluctuation-Induced First-Order Phase
  Transition in an Anisotropic Planar Model of N2 on Graphite}.
\newblock {\em Phys. Rev. Lett.} 48(3):181--184.

\bibitem{Lee1984}
Lee DH, Joannopoulos JD, Negele JW, Landau DP (1984) {Discrete-Symmetry
  Breaking and Novel Critical Phenomena in an Antiferromagnetic Planar XY Model
  in Two Dimensions}.
\newblock {\em Phys. Rev. Lett.} 52(6):433--436.

\bibitem{Fily2014}
Fily Y, Henkes S, Marchetti MC (2014) {Freezing and phase separation of
  self-propelled disks.}
\newblock {\em Soft matter} 10(13):2132--40.

\bibitem{Frenkel1991}
Wojciechowski KW, Frenkel D, Bra\ifmmode~\acute{n}\else \'{n}\fi{}ka AC (1991)
  Nonperiodic solid phase in a two-dimensional hard-dimer system.
\newblock {\em Phys. Rev. Lett.} 66:3168--3171.

\bibitem{Wioland2013}
Wioland H, Woodhouse FG, Dunkel J, Kessler JO, Goldstein RE (2013) {Confinement
  Stabilizes a Bacterial Suspension into a Spiral Vortex}.
\newblock {\em Physical Review Letters} 110(26):268102.

\bibitem{Lushi2014}
Lushi E, Wioland H, Goldstein RE (2014) {Fluid flows created by swimming
  bacteria drive self-organization in confined suspensions}.
\newblock {\em Proc Natl Acad Sci USA} 111(27):9733--9738.

\bibitem{Stark2005}
Stark H, Lubensky TC (2005) {Poisson bracket approach to the dynamics of
  nematic liquid crystals: The role of spin angular momentum}.
\newblock {\em Phys. Rev. E} 72(5):051714.

\bibitem{Scriven1961}
Scriven JDL (1961) {Angular momentum of continua}.
\newblock {\em Nature} 192(4797):36--37.

\bibitem{Furthauer2012}
F\"{u}rthauer S, Strempel M, Grill SW, J\"{u}licher F (2012) {Active chiral
  fluids}.
\newblock {\em Eur Phys J E} 35(9):89.

\bibitem{Manoharan2003}
Manoharan VN, Elsesser MT, Pine DJ (2003) Dense packing and symmetry in small
  clusters of microspheres.
\newblock {\em Science} 301(5632):483--487.

\bibitem{Sacanna2013}
Sacanna S, Pine DJ, Yi GR (2013) {Engineering shape: the novel geometries of
  colloidal self-assembly}.
\newblock {\em Soft Matter} 9:8096--8106.

\bibitem{Ebbens2016}
Ebbens SJ (2016) {Active colloids: Progress and challenges towards realising
  autonomous applications}.
\newblock {\em Curr Opin Colloid Interface Sci} 21:14--23.

\bibitem{Niton2013}
Nitoń P, Żywociński A, Fiałkowski M, Hołyst R (2013) {A
  “nano-windmill” driven by a flux of water vapour: a comparison to the
  rotating ATPase}.
\newblock {\em Nanoscale} 5(20):9732.

\bibitem{Soong2000}
Soong RK et~al. (2000) {Powering an inorganic nanodevice with a biomolecular
  motor.}
\newblock {\em Science (New York, N.Y.)} 290(5496):1555--8.

\bibitem{Maggi2015}
Maggi C, Saglimbeni F, Dipalo M, Angelis FD, Leonardo RD (2015) {Micromotors
  With Asymmetric Shape That Efficiently Convert Light Into Work By
  Thermocapillary Effect}.
\newblock {\em Nat Commun} 6:1--5.

\bibitem{Efrati2014}
Efrati E, Irvine WTM (2014) {Orientation-dependent handedness and chiral
  design}.
\newblock {\em Phys. Rev. X} 4(1):1--12.

\bibitem{Williams2015}
Williams I et~al. (2015) {Transmission of torque at the nanoscale}.
\newblock {\em Nature Physics} 12(1):98--103.

\end{thebibliography}
\end{document}